# Theoretical stiffness limits of 4D printed self-folding metamaterials

Teunis van Manen[1*], Amir A. Zadpoor[1]

[1]*Additive Manufacturing Laboratory, Department of Biomechanical Engineering, Delft University of Technology (TU Delft), Mekelweg 2, Delft 2628CD, The Netherlands.*
[*]*email: t.vanmanen@tudelft.nl*

**ABSTRACT**
4D printing of flat sheets that self-fold into architected 3D structures is a powerful origami-inspired approach for the fabrication of multi-functional devices and metamaterials. The possibility to endow the initially flat sheet with a variety of surface-related functionalities provides the means to simultaneously achieve multi-functional performance and, upon the completion of the self-folding process, a complex 3D architecture. One intrinsic limitation of such a production strategy is the contradictory stiffness requirements for the folding and for subsequent load-bearing steps: while a low stiffness is required to allow for the successful completion of the intended self-folding process, a high stiffness is needed for the subsequent application of the folded structure as a load-bearing mechanical metamaterial. The competition between these two requirements determines the theoretical limits of 4D printed self-folding mechanical metamaterials, which have not been determined before. Here, we present a nonlinear analytical model of self-folding bilayer constructs composed of an active and a passive layer. This finite deformation theoretical model predicts the curvature of activated bilayers, establishes their stability limits, and estimates the stiffness of folded bilayers. Combining the three abovementioned aspects allows for establishing the theoretical stiffness limits of self-folding bilayers. The optimal combinations of geometrical and mechanical properties that result in the highest possible stiffness of folded constructs can then be determined. We compare the predictions of our analytical models with those we obtain computationally as well as the results of our dedicated experiments. Finally, we evaluate the theoretical stiffness limits of bilayer constructs made using the most common types of stimuli-responsive materials. Our analysis shows that a maximum effective modulus of ≈ 1.5 GPa can be achieved using the currently available shape-memory polymers.

**1. INTRODUCTION**
Folding initially flat materials into 3D objects through origami-inspired approaches is a highly promising approach for the fabrication of functionalized metamaterials. One example is the production of 3D hollow polyhedral structures with integrated electronics [1]. Other examples include optical metamaterials made of porous polymer structures patterned with metallic split-ring resonators [2, 3] or tissue engineering scaffolds with different types of (nano)topographical features [4, 5]. What all these examples have in common is that those specific surface features were applied to the surface of an initially flat construct. In this way, planar fabrication techniques (*e.g.*, nanolithography) can be used for small-scale, highly precise ornamentation of the construct surface or for the incorporation of other surface features. Subsequently, self-folding techniques are used for the transformation of the functionalized flat sheets into the desired, complex 3D geometries. Using such an approach, multi-functional, 3D-architected materials can be realized.

Various strategies have been developed for the self-folding of flat sheets into 3D structures [6-8]. The vast majority of such strategies rely on the use of the so-called stimuli responsive materials [8, 9] that exhibit dimensional change upon exposure to a specific stimulus. Two well-known stimuli include temperature that is used for the activation of shape-memory polymers and humidity that induces swelling in hydrogels [6, 10]. A wide range of other responsive materials activated by a variety of different stimuli is reported in the literature [10, 11]. To fabricate self-folding elements, such active materials can be combined with other (*i.e.*, passive) materials to design a specific shape-shifting behavior. A widely applied strategy is to combine two layers of different materials into a bilayer construct [6, 8, 12]. Upon activation, the mismatch in the shape-shifting behaviors of both materials results in the bending of the bilayer specimen. While the fabrication of bilayer constructs can be performed manually, such manual processes are often very laborious and time consuming. The advent of 4D printing processes have made it possible to fabricate bi- and multi-layer constructs fully automatically [13-16]. When combined with rational geometrical design, 4D printing can yield complex shape-shifting behaviors [14, 16] that are otherwise very difficult, if not impossible, to achieve. Moreover, 4D printing can nowadays be performed using inexpensive (*i.e.*, hobbyist) printers and widely available, off-the-shelf, inexpensive materials [16, 17]. 4D printing is, therefore, developing into the *de facto* technique for the production of shape-shifting materials, such as bi-layer constructs.

Depending on their ultimate application, self-folding materials and devices need to satisfy various design requirements. One particularly important requirement, particularly in the context of mechanical metamaterials, is the stiffness of the folded construct. For example, the desired shape of reconfigurable metamaterials needs to be maintained under specific loading conditions to guarantee performance [18-21]. Another example is meta-biomaterials that need to match the mechanical properties of the native bone [4, 5].

The limited stiffness of self-folded constructs is due to an inherent contradiction between the design requirements of the self-folding and load-bearing steps: while a thinner construct with a lower stiffness is required in the self-folding step, a thicker construct is essential for maximizing the load-bearing capacity of the folded structure. To date, not much attention has been paid to these contradictory design requirements and their implications for the stiffness of the resulting metamaterial. Here, we present a nonlinear (*i.e.*, finite deformation hyperelastic) analytical model that can be used to establish the theoretical stiffness limits of self-folded structures and answer the fundamental question: 'given a specific geometrical design and material selection, what is the maximum stiffness that bilayer self-folded structures can achieve?' We demonstrate the validity of our analytical model by comparing its results with those of our dedicated computational models and experiments.

## 2. RESULTS AND DISCUSSION
The starting point of our analysis is a cubic lattice consisting of $n \times n \times n$ unit cells loaded under compression (Figure 1a). Such lattices can be fabricated from arrays of $n \times n \times 1$ unit cells that are folded from initially flat constructs [22] (Figure 1b). The flat sheets are made of an assembly of bilayer elements (red and blue materials in Figure 1) connected by square panels (gray). Driven by the mismatch in the shape-shifting behaviors of the active and passive elements, the panels fold into the programmed 3D geometry upon activation. The initial length, $L$, of the bilayer elements are designed such that folding angles, $θ$, of 90° are achieved (Figure 1c). The other design parameters are the thicknesses of the active and passive layers as well as the material selection for both layers. The width $b$ of the folding elements can be adjusted as well to change the size of the unit cell. The final cubic lattice can be fabricated by stacking the folded arrays either by connecting the individual slices with additional self-folding elements or through the use of other assembly techniques [5, 23-25].

Assuming the panels to be rigid as compared to the bilayer element, the total stiffness of the cubic lattice only depends on the stiffness, $k_{hinge}$, of the folding elements as well as the arrangement of the bilayer elements within the lattice. Considering the lattice depicted in Figure 1a that is made of $n \times n \times n$ unit cells and with the outer dimensions $a \times a \times a$ mm, the effective elastic modulus, $E_L$, of the structure can be determined as follows:

$$E_L = \frac{F}{a^2}\frac{a}{\Delta x} = \frac{1}{a}k_{\text{total}} \quad (1)$$

where $k_{total}$ is the total stiffness of the lattice in N.mm$^{-1}$. The total lattice structure depends on the number of bilayer hinges in series and parallel and equals:

$$k_{\text{total}} = \frac{2n^2}{2n}k_{\text{hinge}} = n\,k_{\text{hinge}} \quad (2)$$

Assuming the hinge width $b$ to equal $a/n$ and by combining Equations 1 and 2, the effective lattice modulus, $E_L$, equals the hinge stiffness normalized by the hinge width, $k_{hinge}/b$. In the remainder of this article, we focus on determining the hinge stiffness of a general bilayer element that folds into a curved beam with a 90° folding angle. The results can, then, be applied to any type of self-folding lattice (of the type shown in Figure 1) using the presented approach.

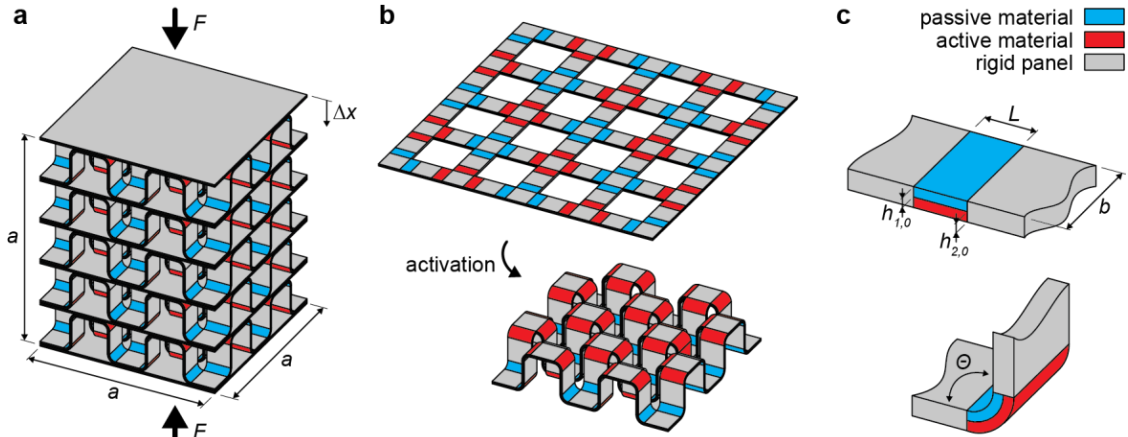

**Figure 1. Self-folding cubic lattice.** (a) A schematic illustration of the compression of a self-folding lattice. (b) Folding of a single array from a flat sheet comprising self-folding bilayer elements (blue and red) and rigid panels (grey). (c) Two rigid panels (grey) connected by a self-folding bilayer element (blue and red).

The analytical model we present here consists of three elements: a finite deformation thick-beam analysis of the bending behavior of bi-layers to relate the geometry, shape shifting response, and mechanical properties of both layers to the curvature of the resulting self-folding construct, a stability model that establishes the limits of stable shape-shifting behavior to avoid wrinkling, and a stiffness prediction model that predicts the stiffness of a self-folded construct given the design parameters of the bilayer. A detailed treatment of all three models and the relevant literature [26-32] is presented in Section 4 (Materials and Methods).

## 2.1. Bilayer bending

Studying the stiffness of self-folded elements requires us to first consider the bending of a general bilayer element. The starting point for our analysis is the plain-strain bending of a thick incompressible hyperelastic beam (Supplementary Figure 1). Following the Rivlin's normality assumptions, the following kinematic relations are imposed [33]:

$$r = f(x) \quad \theta = g(y) \tag{3}$$

Because we are only interested in the final configuration of the activated bilayer, the bending analysis can be decoupled into two separate steps: (a) the free shape-shifting behaviors of the individual layers (b) followed by the bending of the activated layers while enforcing the tie condition at the interface of both layers (Figure 2a). Independent from the underlying shape-shifting mechanism, the following dimensionless parameters can be defined in the activated state:

$$\alpha = \frac{L_2}{L_1} \quad \beta = \frac{h_2}{h_1} \quad \gamma = \frac{G_2}{G_1} \tag{4}$$

where $G_i$ is the shear modulus of layer $i$. Following the imposed kinematics, while adopting an incompressible Neo-Hookean material model for both layers, the static balance equations can be numerically solved. As a result, the normalized curvature of the interface between both layers, $\kappa H$, can be obtained, where $H$ is the combined thickness of the activated layers (*i.e.*, $H = h_1 + h_2$). The normalized bending curvature of the beam is, therefore, a function of only three dimensionless parameters, $\alpha$, $\beta$, and $\gamma$.

## 2.2. Stability analysis

We also performed a stability analysis, using the method of incremental deformation, to determine how far the bilayer element can bend before the appearance of wrinkles on one of its surfaces. In the case of a bilayer assembled from a thin stiff layer and a thick compliant layer, instabilities can occur for small values of $\alpha$ [34]. The results of the bending analysis serve as an input for the stability analysis (see Materials and methods) while a numerical method is employed to detect the bifurcation point.

## 2.3. Stiffness analysis

Based on the described nonlinear bending theory, the initial bilayer dimensions that are required to achieve folding angles of 90° can be determined. The next step is to evaluate the stiffness of the bilayer elements in their folded configuration using the Castigliano's second theorem according to which the small enough displacements resulting from any given force can be calculated as [31, 32]:

$$\delta = \frac{\partial U}{\partial P} \tag{5}$$

where $\delta$ is the displacement of the point of application of force, $P$, along its direction and $U$ is the total strain energy of the elastic system (Figure 3a). The strain energy function of a thick curved bilayer beams is given by the following relationship:

$$U_{\text{thick beam}} = \frac{1}{2}\int_\Theta \frac{M^2}{C_M}d\Theta + \int_\Theta \frac{MN}{C_N}d\Theta + \frac{1}{2}\int_\Theta \frac{N^2 C_C}{C_N^2}d\Theta + \frac{1}{2}\int_\Theta \frac{V^2 C_S}{C_M^2}d\Theta \tag{6}$$

The constants $C_M$, $C_N$, $C_C$, and $C_S$ are provided in the Materials and Methods section. The reaction loads (*i.e.*, normal force $N$, shear force $V$, and moment $M$) at cross-section $\theta$ are a function of the applied load $P$ and centroid axis, $r_c$, of the beam. Differentiation of the strain energy with respect to the applied force $P$ gives the hinge deformation as a function of the applied force $P$, the dimensions $b$, $r_i$, $r_m$, and $r_o$, and the elastic moduli of both layers, $E_i$. The deformation $\delta$ is a linear function of $P$, $b$, and $E_i$. The normalized stiffness $\tilde{k}_{\text{hinge}}$ can, therefore, be calculated as $\tilde{k}_{\text{hinge}} = k_{\text{hinge}} b^{-1} E_2^{-1}$ where $k_{\text{hinge}}$ is the hinge stiffness and equals $P\delta^{-1}$. Normalizing all dimensions with respect to the total thickness of the curved beam $H_c$ reveals that the normalized stiffness $\tilde{k}_{\text{hinge}}$ of a bi-layer quarter-circle beam remains unchanged as the bilayer hinge is scaled and is only a function of the parameters $\alpha$, $\beta_c$, and $\gamma$, where $\beta_c$ is the thickness ratio in the curved configuration.

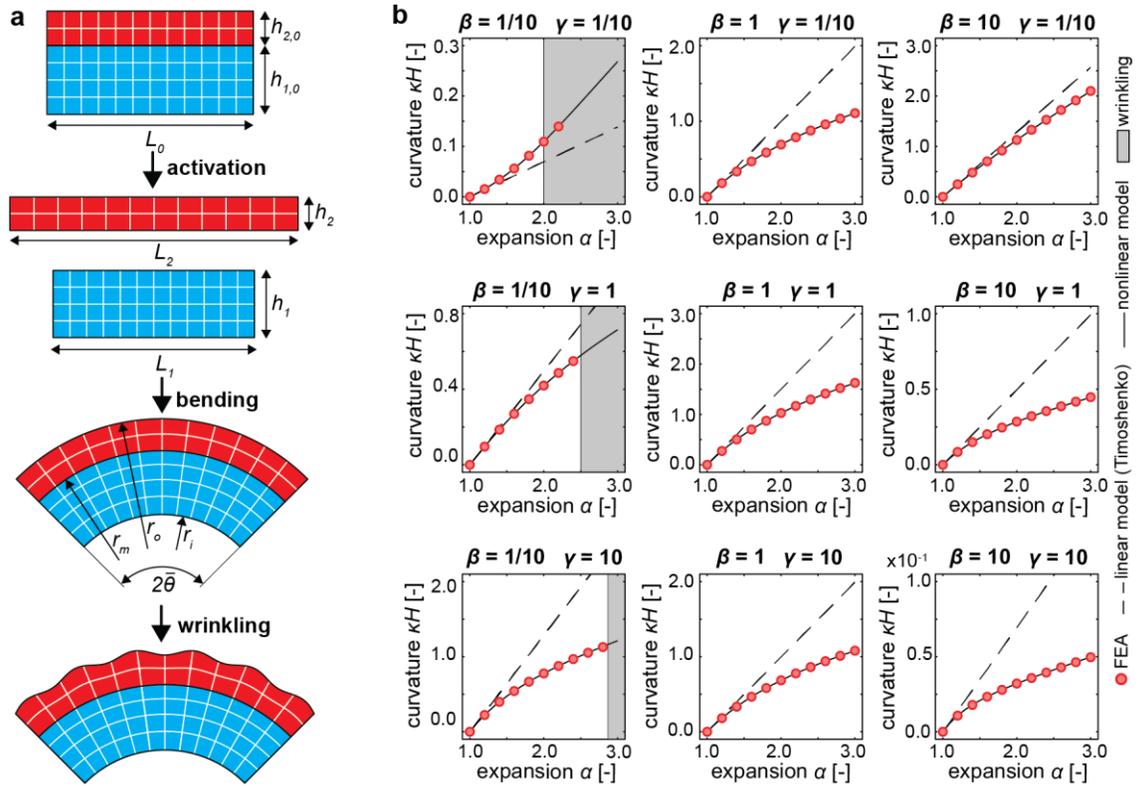

**Figure 2. Bending and wrinkling analysis of self-folding bilayers.** (a) A schematic illustration of the folding and wrinkling of a bilayer element. (b) Theoretical and FEA results for bilayers with varying thickness ratio, $\beta$, and stiffness ratio, $\gamma$.

## 2.4. Computational models

To validate the described theories, a series of finite element analyses (FEA) were performed. First, the results of the plane-strain bending FEA simulations were compared with the theoretical results for the different combinations of $\beta$ and $\gamma$ (Figure 2b). Only the FEA results without the presence of wrinkles are plotted while the gray areas indicate the expansion values for which wrinkling occurs according to the theory. The bending theory is in excellent agreement with FEA while a good agreement between the wrinkling analysis and FEA can be observed as well (Figure 2b). For reference, the predictions of the well-known Timoshenko bilayer bending theory (a linear model) [35] are plotted as well (Figure 2b). Especially for the higher values of $\alpha$, large discrepancies are observed between the linear Timoshenko theory and the nonlinear theory developed here, indicating the need for the described nonlinear theories. Additional simulations were performed to compare the results of the wrinkling analysis with those of the computational models for the different combinations of $\beta$ and $\gamma$. In general, the wrinkling theory is capable of capturing both the onset of wrinkling and the number of wrinkles appearing on the surface of the bilayer (Supplementary Table 1).

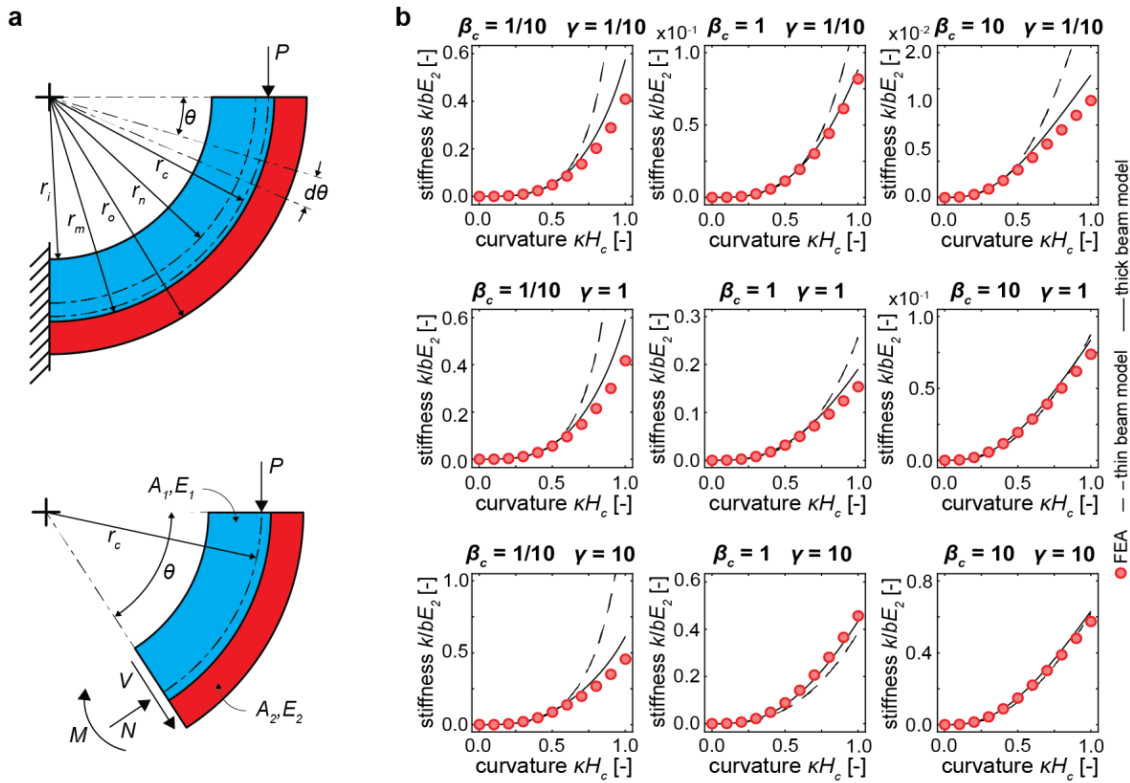

**Figure 3. Stiffness analysis of curved bilayer elements.** (a) A schematic illustration of a folded bilayer specimen. (b) Theoretical and FEA results for bilayers with varying thickness ratio, $\beta_c$, and stiffness ratio, $\gamma$.

Another series of FEA simulations were performed to compare the theoretical stiffness values with the computational results. For small ratios of the initial beam thickness to the radius of the interface, the theories for both thin and thick beams are in good agreement with the FEA results (Figure 3b). While the linear model exhibits very large discrepancies with the computational results when the total beam thickness starts to approach the interface radius, the finite deformation model continues to yield high-fidelity predictions. The advantages of the finite deformation model are particularly clear for the small values of $\beta_c$ where the thin-beam approximation grossly overestimates the stiffness (Figure 3b). This effect is more significant for $\beta_c$ values $<< 1$ because the thickness of the inner layer (*i.e.*, $r_m$-$r_i$) approaches the total beam thickness, meaning that $r_i$ approaches zero for $\kappa H_c$ values close to 1. This is, however, an extreme case for which it is expected that FEA results should deviate from a model that assumes the bilayer elements to behave like a beam. There is, however, a very good agreement between the computational results and the results of the thick-beam model even when the ratio of the initial thickness to the interface radius approaches 1 (Figure 3b).

## 2.5. Experimental validation

As a final validation of both the bending and stiffness theories, we performed several experiments. We 4D printed bilayer bending elements from shape-memory polymers [16] (see Materials and Methods). The bilayers consist of one shrinking layer on top of a semi-passive layer (*i.e.*, a layer with minimal deformation characteristics). For both layers, the longitudinal expansion/shrinkage is coupled to some degree of transverse and vertical deformation. Samples with an expansion ratio $\alpha$ ranging between 1.1 and 2.0 were produced while $\beta$ varied slightly between 0.7 and 1.1. The variation of $\beta$ originates from the fact that samples with equal layer thicknesses were produced resulting in a different layer thickness in the activated states. There was generally a very good agreement between the prediction of the finite deformation model and the experimental results (Figure 4a), demonstrating the ability of the plane strain analytical model to capture the 3D deformation characteristics of actual self-folding specimens.

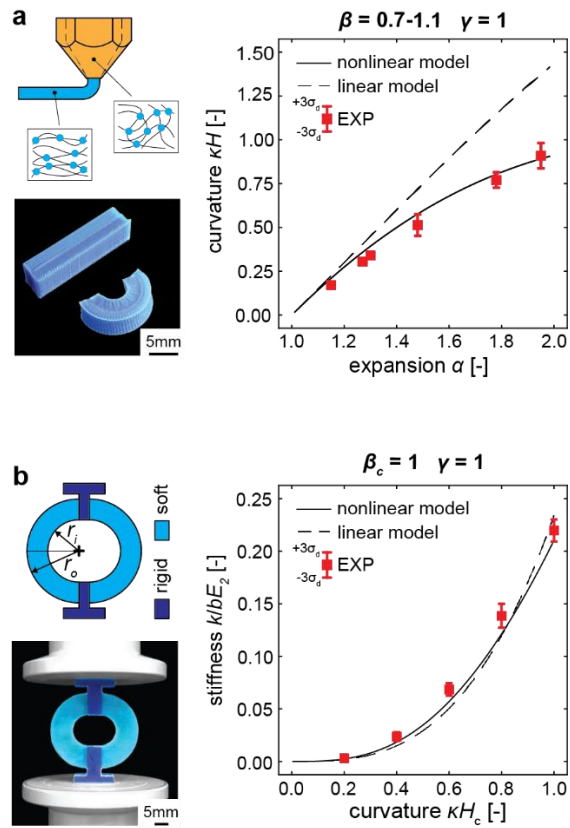

**Figure 4. Experimental validation.** (a) Theoretical and experimental bending results of a self-folding bilayer ($n = 4$). (b) Theoretical and experimental stiffness results of a curved bilayer element ($n = 4$).

To validate the stiffness predictions of the model, various specimens were produced using a multi-material 3D printer working on the basis of UV-curing of multiple types of jetted polymers (the Polyjet technology, see Materials and Methods). In order to perform compressive mechanical testing without a need for additional clamps, the specimens included two parallel sets of soft ($E \approx$ 1.4MPa) quarter-circle beams in series and two rigid clamps (Figure 4b). Because of this arrangement of the beams, the overall sample stiffness equals the stiffness of a single quarter-circle beam. Both layers of the bilayer elements were made from the same material, meaning that the stiffness ratio, $\gamma$, was fixed at 1. The ratio of the interface radius to the total beam thickness was varied between 0.2 and 1.0. Once more, there was a very good agreement between the stiffness values predicted by the finite deformation analytical model and our experimental observations (Figure 4b).

### 2.6. Combined analysis

Combining the three aspects of the analytical model developed here allows for estimating the normalized stiffness of the folded hinges as a function of the three dimensionless parameters $\alpha$, $\beta$, and $\gamma$. For the different values of $\alpha$, the theoretical normalized stiffness can be plotted against the layer thickness ratio, $\beta$, and the stiffness ratio, $\gamma$ (Figure 5a). The results indicate that the stiffness of bilayer hinges can be greatly improved through a proper selection of $\beta$ and $\gamma$. Using $\beta$ and $\gamma$ equal to 1 as the starting point, at least a three-fold increase in the stiffness could be realized by selecting the optimum values of $\beta$ and $\gamma$ (Figure 5a).

Depending on the expansion ratio, $\alpha$, the results of our analysis suggest that the maximum stiffness is achieved for $\beta$ values between 1 and $10^3$ and $\gamma$ values between $10^{-6}$ and $10^{-3}$. In practice, however, it is challenging to realize bilayer specimens with such a large difference in the layer thickness or stiffness. We, therefore, conducted an analysis to search for the maximum stiffness that can be reached while upholding the different bounds limiting the choice of $\beta$ and $\gamma$ values. The upper and lower limits of both $\beta$ and $\gamma$ are defined as $1/C \leq x \leq C$ where $x$ is $\beta$ or $\gamma$ and the boundary $C \geq 1$, respectively. For the different values of $\alpha$, we plotted the maximum stiffness as a function of the limit $C$ (Figure 5b) and found that limiting the maximum stiffness ratio to $10^2$ will be sufficient for realizing at least 80% of the theoretical maximum stiffness (Figure 5c-d). Therefore, $\beta$ values between 2 and 10 (depending on $\alpha$) and a stiffness ratio $\gamma$ of 0.01 would often be a practical choice for the fabrication of bilayer hinges with a near-maximum stiffness.

For small expansion ratios, wrinkling occurs in the bilayers with optimized stiffness (Figure 5a-b). Limiting the feasible values of $\beta$ and $\gamma$ to the stable region only slightly decreases the theoretical maximum stiffness (Figure 5d). It can, therefore, be concluded that optimizing the bilayer design is not significantly hampered by the onset of instabilities.

The stiffness of bilayers with optimized designs were validated using FEA simulations as well. For large values of $\alpha$, the analytical model underestimates the stiffness as compared with FEA (Figure 5b). For stiffness ratios below 0.01, the observed discrepancies are particularly high. It is important to notice that for such a large $\alpha$, the total beam thickness to the interface radius ratio exceeds 1.5, which is beyond the validity limits of a beam-based model. Combined with the effects of a stiffness ratio below $10^{-3}$, the differences between the theoretical model and FEA predictions may even reach 70% (Figure 5b). That said, both theory and FEA exhibit the same overall trends.

Moreover, limiting the stiffness ratio to $10^{-2}$ reduces the discrepancies between both models to a maximum of 40%.

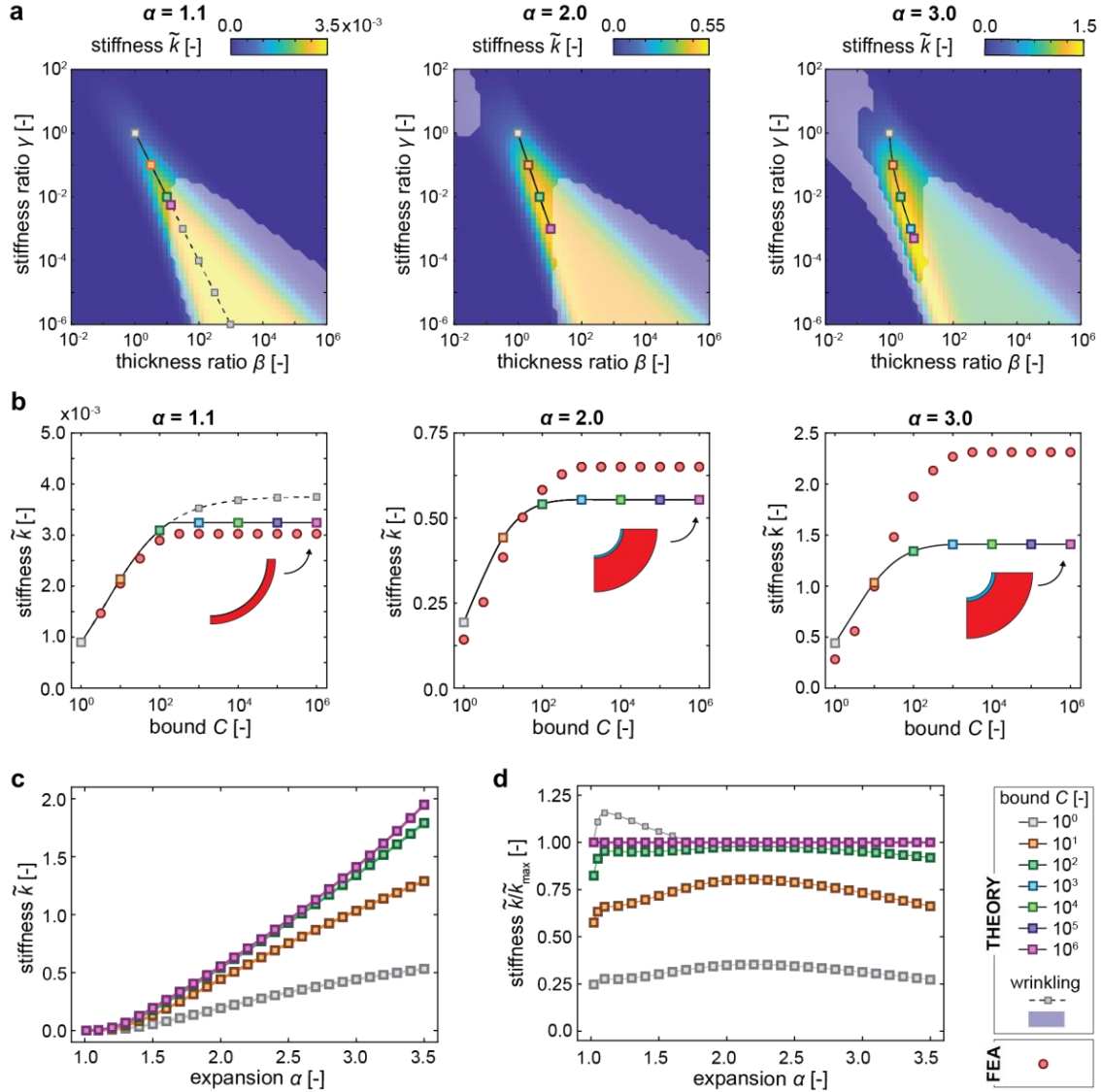

**Figure 5. Combined bending and stiffness analysis.** (a) Theoretical normalized stiffness values for different combinations of thickness ratio, $\beta$, and stiffness ratio, $\gamma$. (b) The maximum theoretical stiffness subjected to different limits for $\beta$ and $\gamma$. (c) The maximum normalized stiffness for different bounds $C$ as a function of the expansion ratio, $\alpha$. (d) The maximum normalized stiffness normalized by the stiffness values with $C = 10^6$.

## 2.7. Practical implications

In the previous section, the bending and stiffness theories were combined to search for the theoretical maximum stiffness of bilayer hinges. For different amounts of expansion, the optimized combinations of $\beta$ and $\gamma$ values were determined. The optimized stiffness of folded bilayer elements only depends on the expansion coefficient, $\alpha$, and the elastic modulus, $E_2$. While it is clear that the combination of a high elastic modulus and a large expansion ratio would result in a high stiffness of folded bilayers, the selection of stimuli-responsive materials is, in practice, a trade-off between the amount of expansion and the material stiffness. Active materials with a high elastic modulus often have small expansion values while the opposite holds true for materials with large expansion coefficients. To estimate the maximum achievable stiffness of bilayer hinges, we need to identify both the elastic modulus and expansion coefficients of the different types of stimuli-responsive materials. Towards that end, we performed a literature study to identify the active materials that are most commonly used for the fabrication of shape-shifting structures. Only studies that reported the fabrication of bilayer folding elements as well as the stiffness and expansion coefficients of the applied materials were included. Four classes of stimuli-responsive materials were identified: i) semiconductor materials with engineered residual stresses (SERS) [36-41], ii) hydrogels [42-50], iii) shape-memory polymers (SMP) [16, 51-58], and iv) liquid crystal elastomers (LCE) [59-65]. The material properties we found for each of those classes of materials are presented in Figure 6a. Assuming that these active materials can be combined with any desired layer of a passive material, the optimized hinge stiffness normalized by the hinge width can be calculated as $k_{\text{hinge}}/b = \tilde{k}_{\text{hinge}} E_2$. The black dashed lines in Figure 6a indicate the constant values of the optimized stiffness $k_{\text{hinge}}/b$.

Comparing the theoretical maximum stiffness values calculated for the different classes of active materials clearly indicates that SMP outperform the other material types (Figure 6b). The combination of moderate expansion coefficients (*i.e.*, $\alpha \sim 2.0$) and elastic moduli up to 2.5 GPa results in a maximum hinge stiffness of $\approx 1.5$ GPa (Figure 6b). Even for $\beta$ and $\gamma$ values equal to 1, the maximum stiffness of SMP bilayer hinges is at least one order of magnitude higher than that of any other optimized bilayer hinge. The bilayers made from SERS or hydrogels display a maximum theoretical stiffness of up to 10 MPa while a somewhat larger stiffness could be achieved using LCE.

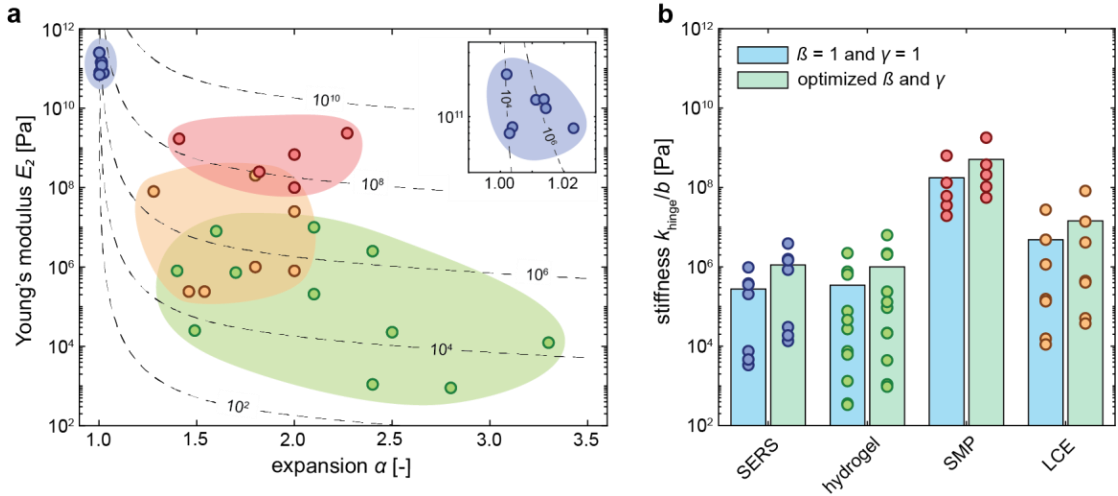

**Figure 6. Stiffness for different active materials.** (a) The elastic moduli and expansion values reported in the literature for the common types of stimuli-responsive materials. The black dashed lines indicate the constant values of $k_{\text{hinge}}/b$. (b) The hinge stiffness of non-optimized and optimized bilayer configurations for the different types of stimuli-responsive materials.

Although these estimated values give a good indication of the achievable hinge stiffness values, some other practical considerations need to be taken into account. First of all, it is assumed that these common types of stimuli-responsive materials could be combined with any sort of passive materials with any desired properties. In the case of SERS materials, however, it is challenging to find a material with a much higher stiffness value. Hence, the optimized stiffness ratios can hardly be realized in practice. One concern regarding SMP is a dramatic drop in their stiffness during their activation [66]. The passive layer should exhibit a similar drop in its stiffness in order to maintain the desired stiffness ratio and realize the theoretically predicted curvature. In practice, this could be achieved by combining the SMP layer with another SMP material. This approach, however, also limits the feasible stiffness ratio. A similar effect can be observed for LCE materials [59, 61, 65]. On the other hand, different types of stimuli-responsive materials could be combined into a bilayer structure to increase the bending curvature. For example, an expanding and a shrinking layer could be combined to create a bilayer with a larger effective expansion ratio, $\alpha$. The combined effects of these practical considerations are highly dependent on the limitations of the specific materials and production techniques and are, thus, difficult to predict.

## 3. OUTLOOK AND CONCLUSIONS

In summary, we presented a finite deformation theory for predicting the bending behavior of bilayer elements, establishing their stability limits, and determining their stiffness. The inputs for the bending analysis are the normalized dimensions of both layers of the composite beam in the activated state that are described by the length ratio $\alpha$, thickness ratio $\beta$, and stiffness ratio $\gamma$. As a result, the normalized radius of the curved bilayer can be calculated, which serves as an input to the stiffness model. For bilayer beams with a 90° folding angle, the normalized stiffness, $\tilde{k}_{\text{hinge}}$, can be obtained as a function of $\alpha$, $\beta$, and $\gamma$. The design of self-folding elements could be optimized in terms of their load-bearing capacity by combining all three aspects of the analytical model. In general, the selection of a thickness ratio $\beta$ between 2 and 10 and a stiffness ratio $\gamma$ of $\approx 10^{-2}$ resulted in a three-fold increase in the hinge stiffness as compared with the default (and naïve) choice of $\beta$ and $\gamma$ values equal to 1. As a final step, the optimized bilayer stiffness values were determined for four types of commonly used stimuli-responsive materials.

The presented theoretical analysis is based on several underlying assumptions. First of all, we performed the analysis based on a cubic self-folding origami-inspired lattice (*e.g.*, similar to the ones presented in [22]). The effective lattice stiffness was found to be equal to the hinge stiffness normalized by the hinge width, $b$. The width $b$ is assumed to be equal to the average width of a unit cell $a/n$. In practice, both the radius and thickness of the folding elements have finite dimensions and the actual hinge width will be smaller than the average unit cell size. A more realistic estimate for the hinge width, $b$, would be possible by subtracting the hinge radius (*i.e.*, $b = a/n - r_i$). The effective modulus of a cubic lattice (Figure 1a) then equals:

$$E_L = \left(1 - \frac{n}{a} r_i\right) \tilde{k}_{\text{hinge}} E_2 \qquad (7)$$

The presented analysis on self-folding bilayers indicates that the stiffness of the bilayer folding elements remains unchanged when all the hinge dimensions are uniformly scaled. A small hinge thickness can, therefore, be selected to reduce the hinge radius, thereby approaching $E_L \approx \tilde{k}_{\text{hinge}} E_2$. It is important to note that for any other arrangements of shape-shifting elements within the self-folding structures, a different relationship between the effective stiffness of the lattice structure and the hinge stiffness may be applicable. In addition to the design of self-folding elements, the arrangement of these elements can be also used as a tool to tune the overall stiffness of self-folding lattices. One example is the use of rectangular elements instead of cubic unit cells to enlarge the effective lattice modulus along one direction.

The finite deformation bending model we presented is based upon several assumption. First, a Neo-Hookean material model was used. Given that the Neo-Hookean material model is only accurate for moderate strain levels, additional FEA simulations were performed using different materials models. A Mooney-Rivlin model, a $2^{nd}$ order Ogden model, and a Yeoh material model were implemented. Comparing the results of those additional models with the theoretical analysis performed using a Neo-Hookean material model indicates that the effects of the selected material model on the predicted response is very limited (Supplementary Figure 2). In addition, the development of a second curvature in the width direction are not captured by our analytical model. Several studies have shown that isotopically expanding bilayers prefer to roll along their longitudinal axis than to develop a double-curved shape [67, 68]. For increasing amounts of expansion (*i.e.*, $α > 1.2$) and large aspect ratios (*i.e.*, $b/L_0 >> 1$), the dominant curvature is found to approach the curvature predicted using the plane-strain theory [68]. This can be explained based on energy considerations as bending is energetically more favorable than the in-plane stretching required for the development of a double curvature. When combined with the effects of clamping, no significant curvature in the width direction is expected.

The performed bifurcation analysis does not predict the wrinkling amplitude but only the onset of wrinkling. The effects of wrinkling on the amount of bending cannot, thus, be determined using the model presented here. While limiting the design space of self-folding bilayers to ensure wrinkling is avoided may, in theory, exclude some high stiffness configurations, the results presented in Figure 5 indicate that the actual drop in the maximum achievable stiffness is in practice quite limited. In addition, this effect is limited to low-expansion values (Figure 5d).

The stiffness of curved bilayer elements was predicted using a model based on a beam theory. In general, however, such models are only accurate for slender structures (*i.e.*, $κH<<1$). Comparing the FEA results with the predictions of the stiffness model indicates that, indeed, larger discrepancies are observed for the larger values of the normalized curvature (Figure 3b). Especially for hydrogels with large expansion coefficients (*i.e.*, $α$~3.0), the normalized curvature $κH$ may even exceed 1 and the beam stiffness will be significantly underestimated (Figure 5b). For the other types of active materials with smaller amounts of expansion (*i.e.*, $α < 2.0$), the discrepancies between the stiffness model and the computational results are < 20% (Figure 5b). A comparison of the stiffness limits for the various types of different active materials shows that the stiffness of hydrogel-based bilayer elements is at least 2 orders of magnitude lower than that of SMP bilayers. Therefore, the underestimation of the stiffness of self-folding hydrogel elements does not affect the overall theoretical stiffness limit of ≈ 1.5 GPa.

Finally, the theoretical stiffness limit of bilayer elements highly depends on the specific properties of the selected stimuli-responsive materials. Here, we have only included the properties of active materials that have been used in the literature for the fabrication of self-folding bilayers. However, other types of stimuli-responsive materials with superior mechanical properties may have been reported in the literature for other applications. Even though the presented list of active materials is not exhaustive, it clearly shows that the application of SMP-based self-folding elements results in significantly higher stiffness values as compared to the other types of activate materials included here.

In conclusion, we developed an analytical model for evaluating the stiffness of bilayer self-folding elements. We then used the developed models to establish the theoretical stiffness limits of self-folding bi-layers. Considering the usual materials used for that purpose, the maximum stiffness reached in practice is estimated to be around 1.5 GPa for SMP-based self-folding elements. The stiffness of SMP bilayers is significantly higher than the stiffness of any other bilayer element based on the other types of active materials. The presented models could be also used for optimizing the stiffness of self-folding bilayers. In general, a thickness ratio between 2 and 10 and a stiffness ratio of around $10^{-2}$ would result in a near-optimal stiffness of the bilayer construct. Given the fact that these results are length scale-independent, they could be applied to various types of self-folding origami lattices.

## 4. MATERIALS AND METHODS
### 4.1. Finite-deformation bending model
We start our analysis by considering the plane strain bending of a thick incompressible beam. The following kinematic relations (normality conditions [33]) are imposed (Supplementary Figure 1):

$$r = f(y) \tag{1.1a}$$
$$\theta = g(x) \tag{1.1b}$$

Based on this assumption, the deformation gradient **F** can be written as follows [28]:

$$\mathbf{F} = \frac{df(y)}{dy} \mathbf{e}_r \otimes \mathbf{e}_y + f(y) \frac{dg(x)}{dx} \mathbf{e}_\theta \otimes \mathbf{e}_x \tag{1.2}$$

Applying the incompressibility condition (*i.e.*, $J = |\mathbf{F}| = 1$) gives:

$$\frac{df(y)}{dy} f(y) \frac{dg(x)}{dx} = 1 \tag{1.3}$$

To solve Equation 1.3, the following boundary conditions can be applied:

$$f\left(y = -\frac{h}{2}\right) = r_i \tag{1.4a}$$
$$f\left(y = \frac{h}{2}\right) = r_o \tag{1.4b}$$
$$g\left(x = -\frac{L}{2}\right) = -\bar{\theta} \tag{1.4c}$$
$$g\left(x = \frac{L}{2}\right) = \bar{\theta} \tag{1.4d}$$

where $r_i$ is the inner radius and $r_o$ is the outer radius (Supplementary Figure 1). By solving Equation 1.3, the principle stretches in the deformed configuration are obtained as [33]:

$$\lambda_r = \frac{dr}{dy} = \frac{df(y)}{dy} = \frac{L_0}{2\bar{\theta}r} \tag{1.5a}$$

$$\lambda_\Theta = r\frac{d\theta}{dx} = f(y)\frac{dg(x)}{dx} = \frac{2\bar{\theta}r}{L_0} \tag{1.5b}$$

As a next step, the balance equations need to be derived. Assuming that the hyperelastic behavior of the involved materials can be described using an incompressible Neo-Hookean material model, the strain energy density function $W$ can be written as:

$$W = \frac{G}{2}(\mathbf{I}_1 - 3) \tag{1.6}$$

where $G$ is the shear modulus and $\mathbf{I}_1$ is the first invariant of the left Cauchy-Green strain tensor, $\mathbf{B} = \mathbf{FF}^T$. The transversal and radial stresses as a function of the principal stretches can be calculated from $\boldsymbol{\sigma} = \frac{1}{J}\frac{\partial W}{\partial \mathbf{F}}\mathbf{F}^T$:

$$\sigma_r = -p + G\lambda_r^2 \tag{1.7a}$$
$$\sigma_\theta = -p + G\lambda_\theta^2 \tag{1.7b}$$

The pressure field $p(r)$ can be obtained up to an integration constant $C$ by substituting Equation 1.7 in the static balance equation ($\nabla \cdot \boldsymbol{\sigma} = 0$) [28]:

$$p(r) = \frac{G}{2}(\lambda_r^2 - \lambda_\theta^2) + C \tag{1.8}$$

Finally, we consider a bilayer constituting of two layers of an incompressible Neo-Hookean material. The volumetric constraint equations ($J = 1$) for both layers yields the following:

$$L_1 h = \bar{\theta}(r_m^2 - r_i^2) \tag{1.9a}$$
$$\alpha L_1 \beta h = \bar{\theta}(r_o^2 - r_m^2) \tag{1.9b}$$

where $r_i$ is the inner radius, $r_m$ is the interface radius, and $r_o$ is the outer radius. Enforcing the zero radial stress condition at the inner and outer radii of the bilayer beam yields the integration constants $C_i$ of both layers. For finding the equilibrium configuration, the net forces and moments must vanish at both ends of the beam:

$$N(\theta = \pm\bar{\theta}) = \int \sigma_\theta dr = \int_{r_1}^{r_2} \sigma_\theta^1 \, dr + \int_{r_2}^{r_3} \sigma_\theta^2 \, dr = 0 \tag{1.10a}$$

$$M(\theta = \pm\bar{\theta}) = \int r \sigma_\theta dr = \int_{r_1}^{r_2} \sigma_\theta^1 \, r \, dr + \int_{r_2}^{r_3} \sigma_\theta^2 \, r \, dr = 0 \tag{1.10b}$$

Equations 1.9 and 1.10 form a set of 4 equations that can be solved numerically to calculate the normalized unknowns $r_i/H$, $r_m/H$, $r_o/H$, and the ratio $\bar{\theta}/L_0$. The normalized results are independent from the initial total thickness, $H$.

## 4.2. Stability model

Using a bending analysis, we derived the static equilibrium shape of a particular self-folding bilayer element. We now aim to find the onset of instability by applying the method of incremental deformations [69]. This method involves the superposition of an incremental displacement field on the bended configuration and calculating the corresponding incremental stresses using [70]:

$$\mathbf{\Sigma} = -\delta p \mathbf{I} + p \mathbf{L}^T + G\mathbf{LB} \quad (2.1)$$

where $\mathbf{\Sigma}$ is the first Piola Kirckhoff stress tensor, $\delta p$ is the pressure increment, $\mathbf{L}$ is the gradient of the incremental displacement field, $\mathbf{u}$ ($\mathbf{L} = \nabla.\mathbf{u}$), and $\mathbf{B} = \mathbf{FF}^T$. Following Roccabianca et al. [29, 30], we propose the following incremental displacement fields:

$$u_r = f(r)\cos(n\theta) \quad (2.2a)$$
$$u_\theta = g(r)\sin(n\theta) \quad (2.2b)$$
$$\delta p = k(r)\cos(n\theta) \quad (2.2c)$$

The incompressibility conditions (i.e., $\text{tr}(\mathbf{L}) = 0$) allows us to write the unknown function $g(r)$ as a function $f(r)$:

$$g = \frac{f + f'r}{n} \quad (2.3)$$

By combining Equations 2.1 and 2.2, the incremental equilibrium equations (i.e., $\nabla.\mathbf{\Sigma} = 0$) can be solved to derive the unknown function $k(r)$ as a function of both $f(r)$ and $g(r)$. The subsequent substitution of Equation 2.3 yields the two following equations [29]:

$$k' = [a^2(1-n^2)]f + \left[a^2r - \frac{2}{a^2r^3}\right]f' + \left[\frac{1}{a^2r^2}\right]f'' \quad (2.4a)$$

$$k = \left[\frac{-1}{a^2r^3}\left(1 - \frac{1}{n^2}\right)\right]f - \left[a^2r^2 + \frac{1}{n^2a^2r^2}\right]f' + \left[\frac{2}{n^2a^2r}\right]f'' + \left[\frac{1}{a^2n^2}\right]f''' \quad (2.4b)$$

where $a = 2\bar{\theta}/L_0$. The differentiation of Equation 2.4b with respect to $r$ and its subsequent substitution into Equation 2.4a results in a single differential equation for the function $f(r)$:

$$f'''' + \left[\frac{2}{r}\right]f''' - \left[n^2a^4r^2 + \frac{3-n^2}{r^2}\right]f'' + \left[\frac{3-n^2}{r^3} + 3n^2a^4r\right]f'$$
$$+ \left[\left(\frac{3}{r^4} + n^2a^4\right)(n^2 - 1)\right]f = 0 \quad (2.5)$$

The function $f$ within each layer can be obtained by enforcing the continuity conditions at the interface of both layers:

$$\Sigma_{rr}^{(1)}(r = r_m) = \Sigma_{rr}^{(2)}(r = r_m) \quad (2.6a)$$
$$\Sigma_{\theta r}^{(1)}(r = r_m) = \Sigma_{\theta r}^{(2)}(r = r_m) \quad (2.6b)$$

$$u_r^{(1)}(r = r_m) = u_r^{(2)}(r = r_m) \tag{2.6c}$$
$$u_\theta^{(1)}(r = r_m) = u_\theta^{(2)}(r = r_m) \tag{2.6d}$$

as well as imposing zero force condition at the inner and outer radii:

$$\Sigma_{rr}^{(1)}(r = r_i) = 0 \tag{2.7a}$$
$$\Sigma_{rr}^{(2)}(r = r_o) = 0 \tag{2.7b}$$
$$\Sigma_{\theta r}^{(1)}(r = r_i) = 0 \tag{2.7c}$$
$$\Sigma_{\theta r}^{(2)}(r = r_o) = 0 \tag{2.7d}$$

The incremental shear forces and normal displacements should also vanish at both ends of the beam:

$$\Sigma_{\theta r}^{(1)}(\theta = \pm\bar{\theta}) = 0 \tag{2.8a}$$
$$\Sigma_{\theta r}^{(2)}(\theta = \pm\bar{\theta}) = 0 \tag{2.8b}$$
$$u_\theta^{(1)} = 0 \tag{2.8c}$$
$$u_\theta^{(2)} = 0 \tag{2.8d}$$

Finally, to avoid numerical ill-conditioning in the cases where large differences between the properties (mechanical, geometrical) of both layers exist, we used the compound matrix method to search for an expansion value $\alpha$ that results in a non-trivial solution of $f^{(i)}$, given the thickness ratio, $\beta$, and stiffness ratio, $\gamma$ [29].

### 4.3. Derivation of the thick-beam stiffness model
Consider a segment of a curved beam loaded in pure bending (Supplementary Figure 3a). We assume that a plane cross-section remains plain after bending. Based on geometrical considerations, the tangential strain at a distance $r - r_n$ from the neutral axis is given by:

$$\epsilon_\theta(r,\theta) = \frac{(r - r_n)d\theta}{r\,\theta} \tag{3.1}$$

Neglecting the radial stresses, the tangential stress for a linear elastic material is:

$$\sigma_\theta(r,\theta) = E\,\epsilon_\theta = E\frac{(r - r_n)d\theta}{r\,\theta} \tag{3.2}$$

In the case of pure bending, the total tangential stress across a specific cross-section equals zero ($\int_A \sigma_\theta\,dA = 0$, where $A$ is the cross-section area). As a result, the neutral axis of a curved beam can be derived as a function from the cross-section geometry and material properties as:

$$r_n = \frac{E_1 A_1 + E_2 A_2}{E_1 \int_{A_1}\frac{1}{r}dA + E_2 \int_{A_2}\frac{1}{r}dA} \tag{3.3}$$

where subscripts 1 and 2 refer to the different layers of the beam. The applied moment $M$ is defined to be positive for increasing curvatures. Combined with Equation 3.2, the moment resulting from the tangential stress is given by:

$$M(\theta) = \int_A (r - r_n)\sigma_\theta \, dA = \frac{d\theta}{\theta}\left[(r_{c,1} - r_n)E_1 A_1 + (r_{c,2} - r_n)E_2 A_2\right] = \frac{d\theta}{\theta} C_M \quad (3.4)$$

Combining Equation 3.4 with Equation 3.2 and adding a term related to the normal force $N$, the tangential stress in layer $i$ can be written as a function of the applied loads as:

$$\sigma_{\theta,i}(r,\theta) = \left(\frac{r - r_n}{r}\right)\frac{ME_i}{C_M} + \frac{NE_i}{C_N} \quad (3.5)$$

where $C_N = E_1 A_1 + E_2 A_2$. The next step is to consider the average transverse shear stress along the width of the beam due to an applied bending moment. Writing the net force in the tangential direction of a beam element with the length $r_c d\theta$ from height $r$ till the outer radius $r_o$ gives (see Supplementary Figure 3b):

$$\tau_2(r,\theta) = \frac{VE_2}{C_M}\left[r_0 - r + r_n \ln\left(\frac{r}{r_0}\right)\right] \quad (3.6)$$

for $r > r_m$. The shear stress in the bottom layer can be obtained in a similar way but by integrating from the inner radius $r_i$ till height $r$. Neglecting the radial stresses, the total internal energy is then obtained as:

$$U = \frac{1}{2}\int_V \frac{\sigma_\theta^2}{E} dV + \frac{1}{2}\int_V \frac{\tau^2}{G} dV \quad (3.7)$$

Substituting Equations 3.5 and 3.6 into Equation 3.7 gives the strain energy as a function of the applied loads:

$$U = \frac{1}{2}\int_\theta \frac{M^2}{C_M} d\theta + \int_\theta \frac{MN}{C_N} d\theta + \frac{1}{2}\int_\theta \frac{N^2 C_C}{C_N^2} d\theta + \frac{1}{2}\int_\theta \frac{V^2 C_S}{C_M^2} d\theta \quad (3.8)$$

where $C_C = r_{c,1} E_1 A_1 + r_{c,2} E_2 A_2$ and $C_s = E_1(1+v)C_{S1} + E_2(1+v)C_{S2}$. The constants $C_{S1}$ and $C_{S2}$ are as follows:

$$C_{s1} = -\frac{r_i^4}{6} + \frac{5r_n r_i^3}{9} - \frac{r_n^2 r_i^2}{2}$$

$$+ \left(r_i^2 r_m^2 - \frac{4r_i r_m^3}{3} + \frac{r_m^4}{2} - r_n r_m^2\left(\left(-2r_i + \frac{4r_m}{3}\right)\ln\left(\frac{r_m}{r_i}\right) + r_i - \frac{4r_m}{9}\right)\right)$$

$$+ r_n^2 r_m^2 \left(\ln\left(\frac{r_m}{r_i}\right)^2 - \ln\left(\frac{r_m}{r_i}\right) + \frac{1}{2}\right)$$

$$C_{s2} = \frac{r_o^4}{6} - \frac{5 r_n r_o^3}{9} + \frac{r_n^2 r_o^2}{2}$$
$$- \left( r_o^2 r_m^2 - \frac{4 r_o r_m^3}{3} + \frac{r_m^4}{2} - r_n r_m^2 \left( \left( -2 r_0 + \frac{4 r_m}{3} \right) \ln\left(\frac{r_m}{r_o}\right) + r_0 - \frac{4 r_m}{9} \right) \right.$$
$$\left. + r_n^2 r_m^2 \left( \ln\left(\frac{r_m}{r_o}\right)^2 - \ln\left(\frac{r_m}{r_o}\right) + \frac{1}{2} \right) \right)$$

where the results of the bending analysis are used to derive explicit relationships for $C_M, C_N, C_C,$ and $C_S$.

For the curved beam shown in Figure 3, the relationships between the applied force $P$ and the cross-sectional loads can be established as:

$$M(\theta) = P\, r_C (1 - \cos(\theta)) \tag{3.10a}$$
$$N(\theta) = P \cos(\theta) \tag{3.10b}$$
$$V(\theta) = -P \sin(\theta) \tag{3.10c}$$

where $r_C$ is the centroid of the composite bilayer beam. The substitution of Equation 3.9 into the strain energy function (Equation 3.8) and subsequent differentiation with respect to the force $P$ ($\delta = \frac{\partial U}{\partial P}$, Castigliano's second theorem) yields the displacement $\delta$ as a linear function of $P$. The initial stiffness $k$ of the curved bilayer can, therefore, be calculated as:

$$k = \frac{P}{\delta} \tag{3.11}$$

### 4.4. Finite element analysis
FEA was performed using the commercial software package Abaqus (Abaqus 6.14, Simulia, US). Static plane strain analyses were performed for both the bending and stiffness steps using an implicit nonlinear solver (Abaqus Standard, full Newton integration). Full-integration hybrid plane strain elements (*i.e.*, CPE4H) were used for modelling the incompressible behavior of the materials. The constitutive behavior of the bilayer during the bending analysis was captured using an incompressible Neo-Hookean material model. In order to allow for instabilities to be triggered, imperfections were introduced through the perturbation of the elastic properties of few elements at the interface of both layers [71, 72]. The expansion of the active layer was modelled through the inclusion of a predefined stress field. The individual stress components were calculated as:

$$\sigma_x = 2 C_{10}^{(2)} \left( \frac{1}{\alpha^2} - \alpha^2 \right) \tag{4.1a}$$
$$\sigma_y = 0 \tag{4.1b}$$
$$\sigma_z = 2 C_{10}^{(2)} (1 - \alpha^2) \tag{4.1c}$$
$$\sigma_{xy} = 0 \tag{4.1d}$$

For the stiffness analysis, an incompressible linear elastic material model was used. A small compressive force was applied and divided by the resulting displacement to calculate the stiffness of the bilayer element.

## 4.5. Bilayer bending experiments

Bilayer samples were fabricated from polylactic acid (PLA) filaments using a fused deposition modeling (FDM) 3D printer (Ultimaker 2+, Ultimaker, The Netherlands). During the process of filament extrusion and the subsequent cooling of the printed material, pre-stresses are stored as memory inside the structure of the printed material [16]. Upon exposure to high temperatures, the printed filament shrinks along the length direction accompanied with expansion in the other directions. Bilayers were produced by printing several layers with filaments aligned in the longitudinal direction on top of a series of layers with filaments perpendicular to the length direction. A nozzle diameter of 0.25 mm and a printing speed of 50 mm/s was used while the layer thickness and extrusion temperature were varied to program the different amounts of longitudinal shrinkage. The specimens were activated through submersion in a transparent container filled with hot water at a temperature of 90 °C for at least 60 s to ensure the completeness of the shape-shifting process. The temperature was controlled by a heating immersion circulator (CORIO CD, Julabo, Germany). The deformed configuration of the activated bilayers was captured using digital cameras. A custom MATLAB (Mathworks, version R2020b, US) code was used for image processing. A circle was fit to the detected interface radius to obtain the curvature of the specimens.

## 4.6. Stiffness experiments

The specimens were produced using a multi-material polyjet 3D printer (Object500 Connex3, Stratasys). The beams were fabricated with a radius of 10.0 mm and a beam thickness varying between 2.0 and 10.0 mm ($\beta$ = 1). The specimens were then mechanically tested under compression (compression rate = 5 mm/min, maximum displacement = 2.0 mm) using a Lloyd LR5K mechanical testing machine equipped with a 100 N load cell. For all the experiments, a linear relationship between the load and displacement could be observed. A MATLAB code was used to fit a line to the experimental data, thereby obtaining the stiffness of the specimens.

**ACKNOWLEDGEMENTS**

The research leading to these results has received funding from the European Research Council under the ERC grant agreement no. [677575].

**REFERENCES**

1. Gracias, D.H., et al., *Forming electrical networks in three dimensions by self-assembly.* science, 2000. **289**(5482): p. 1170-1172.
2. Cho, J.-H., et al., *Nanoscale origami for 3D optics.* Small, 2011. **7**(14): p. 1943-1948.
3. Jamal, M., A.M. Zarafshar, and D.H. Gracias, *Differentially photo-crosslinked polymers enable self-assembling microfluidics.* Nature communications, 2011. **2**(1): p. 1-6.
4. Jamal, M., et al., *Directed growth of fibroblasts into three dimensional micropatterned geometries via self-assembling scaffolds.* Biomaterials, 2010. **31**(7): p. 1683-1690.
5. Janbaz, S., et al., *Origami lattices with free-form surface ornaments.* Science advances, 2017. **3**(11): p. eaao1595.
6. Ionov, L., *Polymeric actuators.* Langmuir, 2015. **31**(18): p. 5015-5024.
7. Liu, Y., J. Genzer, and M.D. Dickey, *"2D or not 2D": Shape-programming polymer sheets.* Progress in Polymer Science, 2016. **52**: p. 79-106.
8. van Manen, T., S. Janbaz, and A.A. Zadpoor, *Programming the shape-shifting of flat soft matter.* Materials Today, 2018. **21**(2): p. 144-163.
9. Zhang, Y., et al., *Printing, folding and assembly methods for forming 3D mesostructures in advanced materials.* Nature Reviews Materials, 2017. **2**(4): p. 1-17.

10. Wei, M., et al., *Stimuli-responsive polymers and their applications.* Polymer Chemistry, 2017. **8**(1): p. 127-143.
11. Liu, F. and M.W. Urban, *Recent advances and challenges in designing stimuli-responsive polymers.* Progress in polymer science, 2010. **35**(1-2): p. 3-23.
12. Gracias, D.H., *Stimuli responsive self-folding using thin polymer films.* Current Opinion in Chemical Engineering, 2013. **2**(1): p. 112-119.
13. Ding, Z., et al., *Direct 4D printing via active composite materials.* Science advances, 2017. **3**(4): p. e1602890.
14. Gladman, A.S., et al., *Biomimetic 4D printing.* Nature materials, 2016. **15**(4): p. 413-418.
15. Kim, Y., et al., *Printing ferromagnetic domains for untethered fast-transforming soft materials.* Nature, 2018. **558**(7709): p. 274-279.
16. van Manen, T., S. Janbaz, and A.A. Zadpoor, *Programming 2D/3D shape-shifting with hobbyist 3D printers.* Materials horizons, 2017. **4**(6): p. 1064-1069.
17. van Manen, T., et al., *4D printing of reconfigurable metamaterials and devices.* Communications Materials, 2021. **2**(1): p. 1-8.
18. Filipov, E.T., T. Tachi, and G.H. Paulino, *Origami tubes assembled into stiff, yet reconfigurable structures and metamaterials.* Proceedings of the National Academy of Sciences, 2015. **112**(40): p. 12321-12326.
19. Overvelde, J.T., et al., *Rational design of reconfigurable prismatic architected materials.* Nature, 2017. **541**(7637): p. 347-352.
20. Zhai, Z., Y. Wang, and H. Jiang, *Origami-inspired, on-demand deployable and collapsible mechanical metamaterials with tunable stiffness.* Proceedings of the National Academy of Sciences, 2018. **115**(9): p. 2032-2037.
21. Bertoldi, K., et al., *Flexible mechanical metamaterials.* Nature Reviews Materials, 2017. **2**(11): p. 1-11.
22. Nojima, T. and K. Saito, *Development of newly designed ultra-light core structures.* JSME International Journal Series A Solid Mechanics and Material Engineering, 2006. **49**(1): p. 38-42.
23. van Manen, T., et al., *Kirigami-enabled self-folding origami.* Materials Today, 2020. **32**: p. 59-67.
24. Randhawa, J.S., et al., *Importance of surface patterns for defect mitigation in three-dimensional self-assembly.* Langmuir, 2010. **26**(15): p. 12534-12539.
25. Tien, J., T.L. Breen, and G.M. Whitesides, *Crystallization of millimeter-scale objects with use of capillary forces.* Journal of the American Chemical Society, 1998. **120**(48): p. 12670-12671.
26. Abdolahi, J., et al., *Analytical and numerical analysis of swelling-induced large bending of thermally-activated hydrogel bilayers.* International Journal of Solids and Structures, 2016. **99**: p. 1-11.
27. Morimoto, T. and F. Ashida, *Temperature-responsive bending of a bilayer gel.* International Journal of Solids and Structures, 2015. **56**: p. 20-28.
28. Nardinocchi, P. and E. Puntel, *Finite bending solutions for layered gel beams.* International Journal of Solids and Structures, 2016. **90**: p. 228-235.
29. Roccabianca, S., D. Bigoni, and M. Gei, *Long wavelength bifurcations and multiple neutral axes of elastic layered structures subject to finite bending.* Journal of Mechanics of Materials and Structures, 2011. **6**(1): p. 511-527.
30. Roccabianca, S., M. Gei, and D. Bigoni, *Plane strain bifurcations of elastic layered structures subject to finite bending: theory versus experiments.* IMA journal of applied mathematics, 2010. **75**(4): p. 525-548.
31. Budynas, R.G., *Advanced strength and applied stress analysis*. 1999: McGraw-Hill.
32. Budynas, R.G., J.K. Nisbett, and K. Tangchaichit, *Shigley's mechanical engineering design*. 2005: McGraw Hill New York.
33. Rivlin, R., *Large elastic deformations of isotropic materials. V. The problem of flexure.* Proceedings of the Royal Society of London. Series A. Mathematical and Physical Sciences, 1949. **195**(1043): p. 463-473.

34. Cendula, P., et al., *Bending and wrinkling as competing relaxation pathways for strained free-hanging films.* Physical Review B, 2009. **79**(8): p. 085429.
35. Timoshenko, S., *Analysis of bi-metal thermostats.* Josa, 1925. **11**(3): p. 233-255.
36. Bassik, N., et al., *Patterning thin film mechanical properties to drive assembly of complex 3D structures.* Advanced Materials, 2008. **20**(24): p. 4760-4764.
37. Arora, W.J., et al., *Membrane folding to achieve three-dimensional nanostructures: Nanopatterned silicon nitride folded with stressed chromium hinges.* Applied physics letters, 2006. **88**(5): p. 053108.
38. Pinto, R.M., V. Chu, and J.P. Conde, *Amorphous Silicon Self-Rolling Micro Electromechanical Systems: From Residual Stress Control to Complex 3D Structures.* Advanced Engineering Materials, 2019. **21**(9): p. 1900663.
39. Schmidt, O., et al., *Self-assembled nanoholes, lateral quantum-dot molecules, and rolled-up nanotubes.* IEEE Journal of selected topics in quantum electronics, 2002. **8**(5): p. 1025-1034.
40. Vaccaro, P.O., K. Kubota, and T. Aida, *Strain-driven self-positioning of micromachined structures.* Applied Physics Letters, 2001. **78**(19): p. 2852-2854.
41. Pi, C.-H. and K.T. Turner, *Design, analysis, and characterization of stress-engineered 3D microstructures comprised of PECVD silicon oxide and nitride.* Journal of Micromechanics and Microengineering, 2016. **26**(6): p. 065010.
42. Cheng, X., et al., *Surface chemical and mechanical properties of plasma-polymerized N-isopropylacrylamide.* Langmuir, 2005. **21**(17): p. 7833-7841.
43. Bassik, N., et al., *Photolithographically patterned smart hydrogel based bilayer actuators.* Polymer, 2010. **51**(26): p. 6093-6098.
44. Kim, J., et al., *Reversible self-bending soft hydrogel microstructures with mechanically optimized designs.* Chemical Engineering Journal, 2017. **321**: p. 384-393.
45. Naficy, S., et al., *4D printing of reversible shape morphing hydrogel structures.* Macromolecular Materials and Engineering, 2017. **302**(1): p. 1600212.
46. Na, J.H., et al., *Programming reversibly self-folding origami with micropatterned photo-crosslinkable polymer trilayers.* Advanced Materials, 2015. **27**(1): p. 79-85.
47. Hu, Z., X. Zhang, and Y. Li, *Synthesis and application of modulated polymer gels.* Science, 1995. **269**(5223): p. 525-527.
48. Morales, D., et al., *Ionoprinted multi-responsive hydrogel actuators.* Micromachines, 2016. **7**(6): p. 98.
49. Wang, X., et al., *Multi-Responsive Bilayer Hydrogel Actuators with Programmable and Precisely Tunable Motions.* Macromolecular Chemistry and Physics, 2019. **220**(6): p. 1800562.
50. Jamal, M., et al., *Bio-origami hydrogel scaffolds composed of photocrosslinked PEG bilayers.* Advanced healthcare materials, 2013. **2**(8): p. 1142-1150.
51. Janbaz, S., R. Hedayati, and A. Zadpoor, *Programming the shape-shifting of flat soft matter: from self-rolling/self-twisting materials to self-folding origami.* Materials Horizons, 2016. **3**(6): p. 536-547.
52. Felton, S.M., et al., *Self-folding with shape memory composites.* Soft Matter, 2013. **9**(32): p. 7688-7694.
53. Ge, Q., H.J. Qi, and M.L. Dunn, *Active materials by four-dimension printing.* Applied Physics Letters, 2013. **103**(13): p. 131901.
54. Ge, Q., et al., *Active origami by 4D printing.* Smart materials and structures, 2014. **23**(9): p. 094007.
55. Liu, Y., et al., *Self-folding of polymer sheets using local light absorption.* Soft matter, 2012. **8**(6): p. 1764-1769.
56. Mailen, R.W., et al., *Modelling of shape memory polymer sheets that self-fold in response to localized heating.* Soft Matter, 2015. **11**(39): p. 7827-7834.
57. Wang, H., et al., *Shape-Controlled, Self-Wrapped Carbon Nanotube 3D Electronics.* Advanced Science, 2015. **2**(9): p. 1500103.

58. Liu, Y., et al., *Sequential self-folding of polymer sheets.* Science Advances, 2017. **3**(3): p. e1602417.
59. Shaha, R.K., A.H. Torbati, and C.P. Frick, *Body-temperature s hape-shifting liquid crystal elastomers.* Journal of Applied Polymer Science, 2021. **138**(14): p. 50136.
60. An, N., M. Li, and J. Zhou, *Instability of liquid crystal elastomers.* Smart Materials and Structures, 2015. **25**(1): p. 015016.
61. Kim, H., et al., *Tough, shape-changing materials: crystallized liquid crystal elastomers.* Macromolecules, 2017. **50**(11): p. 4267-4275.
62. Agrawal, A., et al., *Shape-responsive liquid crystal elastomer bilayers.* Soft Matter, 2014. **10**(9): p. 1411-1415.
63. Barnes, M. and R. Verduzco, *Direct shape programming of liquid crystal elastomers.* Soft Matter, 2019. **15**: p. 870-879.
64. Boothby, J. and T. Ware, *Dual-responsive, shape-switching bilayers enabled by liquid crystal elastomers.* Soft Matter, 2017. **13**(24): p. 4349-4356.
65. Yuan, C., et al., *3D printed reversible shape changing soft actuators assisted by liquid crystal elastomers.* Soft Matter, 2017. **13**(33): p. 5558-5568.
66. Meng, H. and G. Li, *A review of stimuli-responsive shape memory polymer composites.* Polymer, 2013. **54**(9): p. 2199-2221.
67. Alben, S., B. Balakrisnan, and E. Smela, *Edge Effects Determine the Direction of Bilayer Bending.* Nano Letters, 2011. **11**(6): p. 2280-2285.
68. Finot, M. and S. Suresh, *Small and large deformation of thick and thin-film multi-layers: effects of layer geometry, plasticity and compositional gradients.* Journal of the Mechanics and Physics of Solids, 1996. **44**(5): p. 683-721.
69. Ogden, R.W., *Non-linear elastic deformations*. 1997: Courier Corporation.
70. Su, Y., et al., *Pattern evolution in bending dielectric-elastomeric bilayers.* Journal of the Mechanics and Physics of Solids, 2020. **136**: p. 103670.
71. Nikravesh, S., D. Ryu, and Y.-L. Shen, *Direct numerical simulation of buckling instability of thin films on a compliant substrate.* Advances in Mechanical Engineering, 2019. **11**(4): p. 1687814019840470.
72. Nikravesh, S., D. Ryu, and Y.-L. Shen, *instabilities of thin films on a compliant Substrate: Direct numerical Simulations from Surface Wrinkling to Global Buckling.* Scientific reports, 2020. **10**(1): p. 1-19.

Supplementary document to
# Theoretical stiffness limits of 4D printed self-folding metamaterials

Teunis van Manen[1*], Amir A. Zadpoor[1]

[1]Additive Manufacturing Laboratory, Department of Biomechanical Engineering, Delft University of Technology (TU Delft), Mekelweg 2, Delft 2628CD, The Netherlands.
*email: t.vanmanen@tudelft.nl

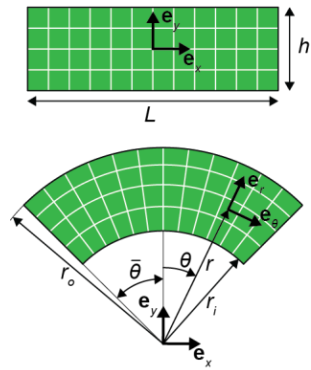

**Supplementary Figure 1.** Schematic representation of the bending of a thick elastic block into a curved beam.

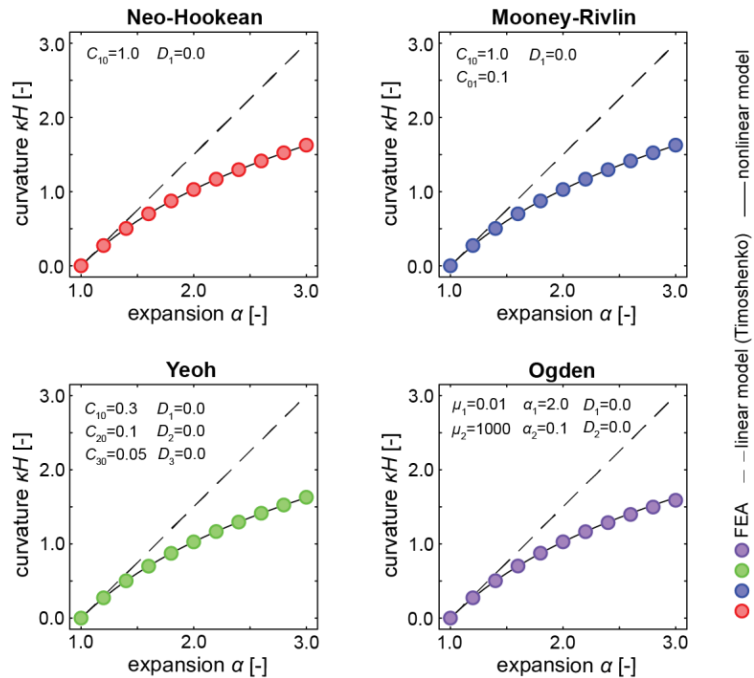

**Supplementary Figure 2.** Theoretical and FEA results for the bending of bilayer beams using different incompressible hyperelastic material models.

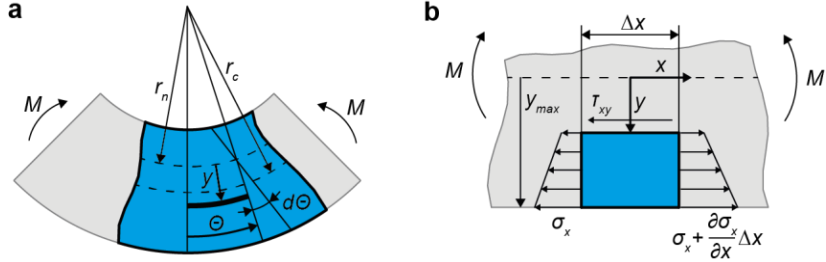

**Supplementary Figure 3. Schematic illustrations of beam segments.** (a) A curved beam loaded in pure bending. (b) A straight beam loaded in bending. The net force in x-direction over the beam element in blue equals zero.

**Supplementary Table 1.** Comparison of the wrinkling theory and FEA for multiple combinations of thickness ratio and stiffness ratio. Both the expansion values at the onset of wrinkling $α_{wrinkling}$ and the number of wrinkles $n_{wrinkles}$ are listed while the maximum value of $α$ was set to 3.0.

|   |   | **Theory** |   | **FEA** |   |
|---|---|---|---|---|---|
| $β$ | $γ$ | $α_{wrinkling}$ | $n_{wrinkles}$ | $α_{wrinkling}$ | $n_{wrinkles}$ |
| 1/100 | 1/100 | >3 | n/a | >3 | n/a |
| 1/100 | 1/10 | >3 | n/a | >3 | n/a |
| 1/100 | 1 | 1.9 | 151 | 2.0 | 148 |
| 1/100 | 10 | 1.2 | 134 | 1.3 | 85 |
| 1/100 | 100 | 1.2 | 22 | 1.2 | 22 |
| 1/10 | 1/100 | 3.0 | 170 | 2.1 | 140 |
| 1/10 | 1/10 | 2.1 | 88 | 2.3 | 70 |
| 1/10 | 1 | 2.6 | 6 | 2.5 | 7 |
| 1/10 | 10 | 2.9 | 2 | 2.7 | 2 |
| 1/10 | 100 | 2.6 | 2 | 2.9 | 2 |
| 1 | 1/100 | >3 | n/a | >3 | n/a |
| 1 | 1/10 | >3 | n/a | >3 | n/a |
| 1 | 1 | >3 | n/a | >3 | n/a |
| 1 | 10 | >3 | n/a | >3 | n/a |
| 1 | 100 | >3 | n/a | >3 | n/a |
| 10 | 1/100 | 1.05 | 2 | 1.1 | 2 |
| 10 | 1/10 | >3 | n/a | >3 | n/a |
| 10 | 1 | >3 | n/a | >3 | n/a |
| 10 | 10 | >3 | n/a | >3 | n/a |
| 10 | 100 | >3 | n/a | >3 | n/a |
| 100 | 1/100 | 1.03 | 2 | 1.07 | 2 |
| 100 | 1/10 | >3 | n/a | >3 | n/a |
| 100 | 1 | >3 | n/a | >3 | n/a |
| 100 | 10 | >3 | n/a | >3 | n/a |
| 100 | 100 | >3 | n/a | >3 | n/a |